\documentclass[pra,twocolumn,notitlepage,superscriptaddress,showpacs, amsmath, amssymb]{revtex4-2}

\usepackage[english]{babel}

\usepackage[T1]{fontenc}
\usepackage{textcomp}
\usepackage{amsmath,amssymb}
\usepackage{braket}
\usepackage{color, xcolor}
\usepackage{graphicx} 
\DeclareGraphicsExtensions{{.pdf}}
\usepackage{hyperref}
\usepackage{soul}
\usepackage{lipsum}
\usepackage{booktabs}
\usepackage{color}

\definecolor{lccolor}{rgb}{0.0, 0.3, 0.7}

\begin{document}
\title{Controlling Atom Array in an Ultra-high-cooperativity Optical Cavity}
\author{Jilai Ye}
\altaffiliation{These authors contributed equally to this work.}
\author{Zhihao Chi}
\altaffiliation{These authors contributed equally to this work.}
\author{Ye Tian}
\altaffiliation{These authors contributed equally to this work.}
\author{Shuyao Mei}
\author{Xiaoyu Li}
\author{Wanlin Wang}
\author{Wenjun Zhang}
\author{Yajuan Zhao}
\affiliation{Department of Physics and State Key Laboratory of Low Dimensional Quantum Physics, Tsinghua University, Beijing, 100084, China}
\author{Jiazhong Hu}
\email{hujiazhong01@ultracold.cn}
\affiliation{Quantum Science Center of Guangdong-Hong Kong-Macao Greater Bay Area, Shenzhen 518045, China}
\author{Wenlan Chen}
\email{cwlaser@mail.tsinghua.edu.cn}
\affiliation{Department of Physics and State Key Laboratory of Low Dimensional Quantum Physics, Tsinghua University, Beijing, 100084, China}
\affiliation{Beijing Academy of Quantum Information Sciences, Beijing, 100193, China}
\affiliation{Beijing Key Laboratory of Cold Atom Quantum Computation, Tsinghua University, Beijing, 100084, China}
\begin{abstract}
Neutral-atom array and cavity quantum electrodynamics offer complementary strengths for quantum science: scalable, reconfigurable qubit architectures and strong coherent light-matter coupling. Combining them in a single platform requires an optical cavity with simultaneously high cooperativity, sufficient mode volume to accommodate atom array, and ample side optical access for atom trapping, imaging, cooling, and rearrangement, a combination that is challenging to achieve. 
Here we realize an atomic array integrated with a millimeter-scale Fabry--P\'erot cavity whose optically-characterized single-atom cooperativity reaches $\eta_{\mathrm{cav}}=125\pm13$. Atom-cavity transmission spectra of trapped atoms yield an effective spectroscopic cooperativity $\eta_{\mathrm{spec}}=106\pm 3$, providing an in-situ verification of strong coupling in the integrated platform. In this integrated system, we achieve a state-measurement fidelity of $99.925^{+15}_{-18}\%$ for a single atom with a 4-$\mu$s readout time. We also demonstrate simultaneous coupling of up to 21 individually trapped atoms to the antinode of the cavity mode.
The key technical advance is a two-step mirror-fabrication method combining precision mechanical shaping and carbon-dioxide laser polishing, which produces concave fused-silica mirrors with sub-millimeter radii of curvature and residual roughness below 2~\AA. Our results establish a regime of cavity-integrated atomic array that simultaneously provides high cooperativity, large mode volume, and flexible manipulation of individual atoms, opening opportunities for cavity-assisted quantum state readout and long-range entanglement-engineering in atom-array platforms.

\end{abstract}

\maketitle

Neutral-atom array and cavity quantum electrodynamics (QED) provide powerful but historically distinct routes toward quantum science. Optical tweezer array offer scalable and reconfigurable architectures with high-fidelity local control \cite{Barredo2016,Endres2016,Browaeys2020,Kaufman2021,Saffman2010}, while cavity QED enables strong atom-photon coupling, long-range coherent interactions in atoms, and efficient quantum state readout \cite{Kimble1998,Miller2005,Reiserer2015,Reiserer2022,Brennecke2007}. Integrating both capabilities in a single platform would enable neutral-atom systems with enhanced connectivity, cavity-assisted quantum state measurement, and collective quantum control \cite{Deist2022PRL,Yan2023PRL,Hu2025PRL,Grinkemeyer2025Science,Wang2025QuantumFrontiers,Zhang2024PRR,Shaw2026Nature,Chen2022OptExpress,Huie2021PRR,Covey2023npjQI,Seubert2025PRXQuantum,Hartung2024Science,de2026realization,PhysRevLett173601,Benjamin2025}.

\begin{figure}[t]
    \centering
    \includegraphics[width=0.85\linewidth]{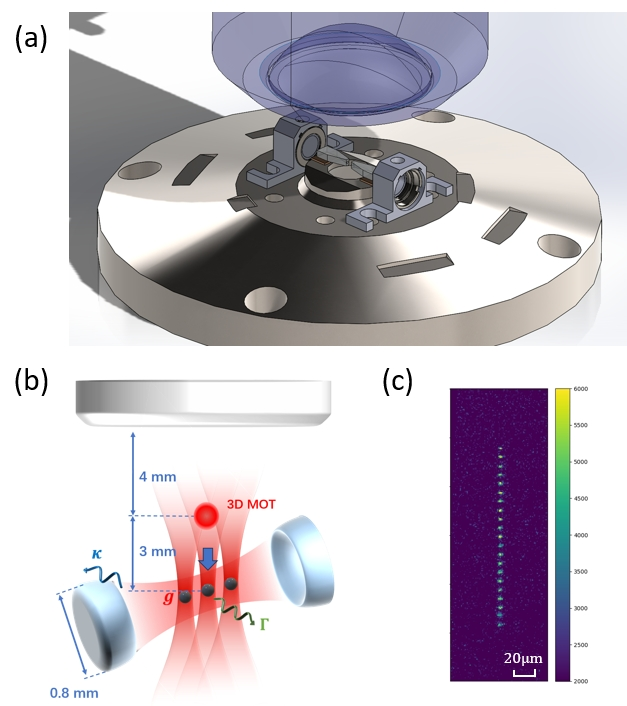}
    \caption{Integrated cavity-atom array platform. (a) The mechanical design of the cavity assembly, in which two concave mirrors on top of the shear piezo actuators, and collimation lenses are mounted on titanium structures. (b) Schematic of the optical tweezer array coupled to the cavity mode, with low-aberration viewport above them. The coherent atom-cavity coupling rate $g$, cavity decay rate $\kappa$, and atomic linewidth $\Gamma$ are indicated. (c) Fluorescence image of atoms loaded into tweezers overlapping with the cavity mode. In the present implementation, 21 individually trapped atoms are rearranged and loaded into the cavity mode and coupled to the antinode of the cavity simultaneously.}
    \label{fig:platform}
\end{figure}

\begin{figure}
    \centering
    \includegraphics[width=0.95\linewidth]{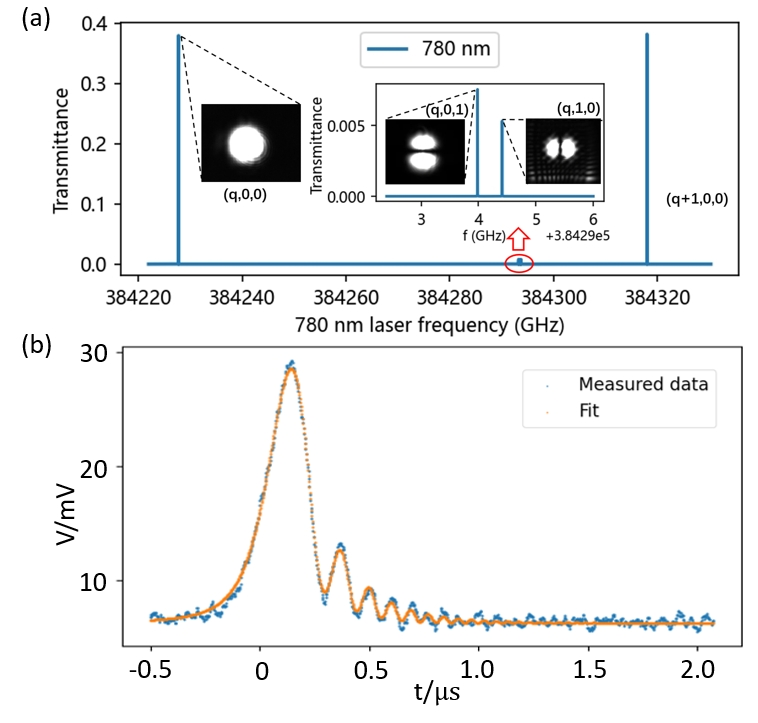}
    \caption{Cavity characterization. (a) Transmission spectrum of a $780 ~\mathrm{nm}$ probe laser sweeping over one free spectral range (FSR) of the cavity. The frequency spacing between two adjacent fundamental transverse ($\mathrm{TEM}_{00}$) modes is the $\nu_{\mathrm{FSR}}=89.7 ~ \mathrm{GHz}$, which gives a cavity length $L=1.67~\mathrm{mm}$. In the cavity mode labeling (q, m, n), q is the longitudinal mode index, m and n are the transverse mode indices. The frequency spacings between the $\mathrm{TEM}_{01}$ ($\mathrm{TEM}_{10}$) mode and  $\mathrm{TEM}_{00}$ mode are $66.0 ~ \mathrm{GHz}$ ($66.5 ~ \mathrm{GHz}$). The inset shows the transmission spectrum of $\mathrm{TEM}_{01}$ and $\mathrm{TEM}_{10}$ modes. (b) Ring-down measurement of the cavity. This is achieved by rapidly sweeping the laser frequency across the resonance while using linearly polarized incident light to avoid polarization splitting. The fit yields a cavity linewidth $\kappa=0.94 \pm 0.09~\mathrm{MHz}$ .}
    \label{fig:Opticalmeasurement}
\end{figure}

\begin{figure*}[t]
    \centering
    \includegraphics[width=0.95\linewidth]{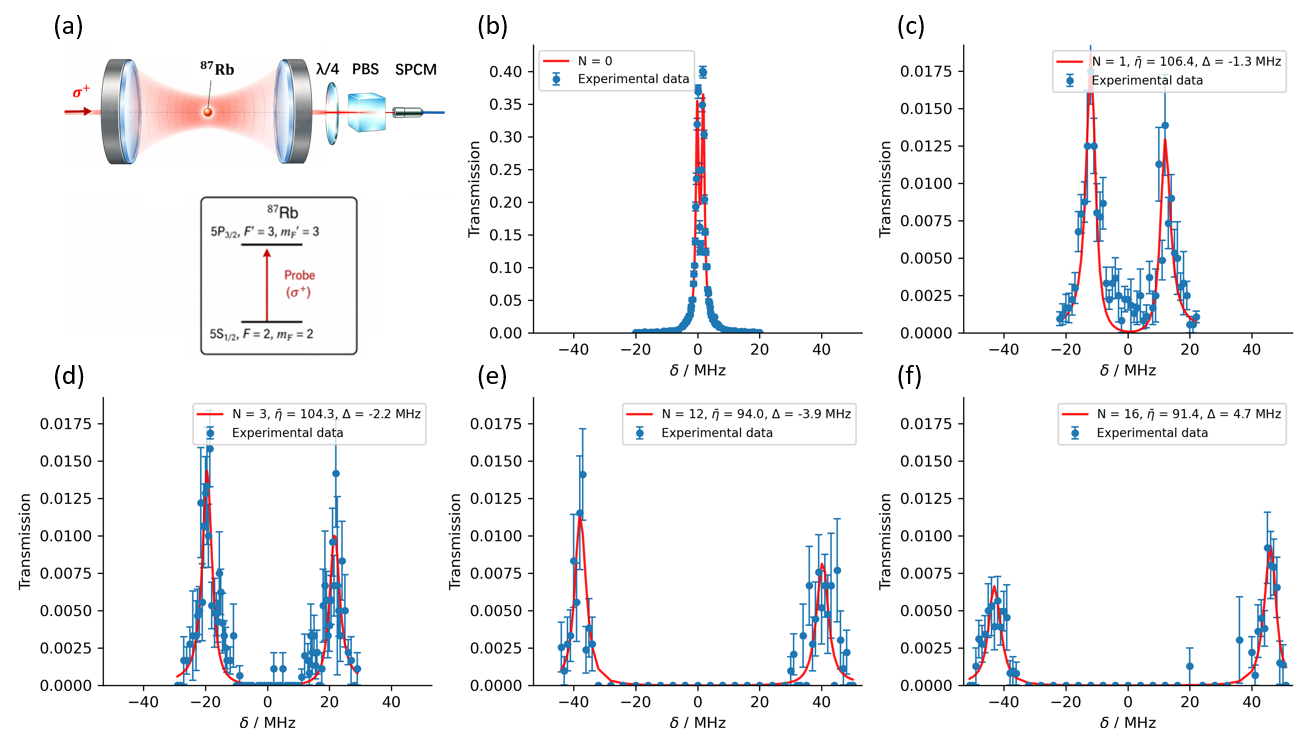}
    \caption{Atom-cavity transmission spectra. (a) Transmission spectra measurment scheme and the atomic transition used in the spectroscopy, $5S_{1/2}|F=2,m_F=2\rangle \rightarrow 5P_{3/2}|F'=3,m_{F'}=3\rangle$ of $^{87}$Rb. (b) Empty cavity transmission spectrum. Here, $\delta=\omega-\omega_c$ is the frequency difference between the probe light $\omega$ and the cavity resonance $\omega_c$, and $\Delta=\omega_a-\omega_c$ is the atom-cavity detuning. The observed doublet arises from the fact that the two nearly orthogonal cavity polarization eigenmodes have slightly different resonance frequencies. (c)(d)(e)(f) Spectra for N=1,3,12,16 atoms trapped at the antinode inside the cavity mode.  Blue points are experimental data and red curves are fits to a cavity-QED model that includes the spatial distribution of atoms thermally broadened by an temperature of 13$~\mathrm{\mu K}$. The fitting yields the average single-atom cooperativity $\eta_{\text{spec}}$ for each atom number: $106\pm 3$ ($N=1$), $104.3 \pm 1.5$ ($N=3$), $94.0 \pm 1.2$ ($N=12$), and $91.4 \pm 0.8$ ($N=16$).
    }    
    %The fits yield a single-atom cooperativity of $\eta = 110\pm10$.}
    \label{fig:spectra}
\end{figure*}

A core challenge lies in the conflicting technical requirements between the two platforms. High-cooperativity cavities, in which the cooperativity parameter $\eta = \frac{4g^2}{\kappa \Gamma}$ quantifies the ratio of coherent interaction strength to dissipation for intracavity atoms \cite{TANJISUZUKI2011201}, feature strong optical mode confinement and low optical loss. Such cavities are typically realized using short microcavities characterized by small mode volumes and limited optical access.
By contrast, trapping atom array require millimeter-scale geometries, large mode volume, and high-numerical-aperture access from cavity side for trapping, imaging, cooling, and rearrangement \cite{Barredo2016,Endres2016,Browaeys2020,Kaufman2021}. In order to cooperate capabilities of both platforms, scientists tried different schemes, including bow-tie cavity to simultaneously enable long cavity length and small mode waist\cite{peters2025cavity,Zhang2024PRR}, fiber cavity with careful engineering of side optical access for cavity tweezer array\cite{Hunger2010,roschinski2026,picot2026,Grinkemeyer2025Science,PhysRevLett240802}, and lens array in the cavity focusing the optical mode waist to $\sim$1~$\mu$m to enable both  high-cooperativity and optical tweezers via the focused beams \cite{Shaw2026Nature}.

Here, we build a millimeter-scale Fabry-Perot cavity with single-atom cooperativity $\eta_{\mathrm{cav}}=125\pm13$, capable to contain up to 200 individually controlled atoms in the atom array. The key element of this system is a pair of superpolished mirrors with radius of curvature of one millimeter. It is fabricated using a two-step method that combines precision mechanical shaping with controlled carbon-dioxide laser polishing \cite{Petrak2011,Hunger2012AIP}. This approach produces concave fused-silica mirrors with sub-millimeter radii of curvature and ultra-low surface roughness, thereby enabling high cavity finesse of 95,000, small cavity mode waist of 9.5~$\mu$m, while maintaining a compatible geometry with convenient optical tweezers operations from the side of the cavity.

Fig.~\ref{fig:platform} shows the illustration of the integrated platform. The Fabry-Perot cavity has a length of $L=1.67$~mm and is built with two in-house-fabricated concave mirrors. Each mirror is mounted on a shear piezo actuator using low-outgassing epoxy (EPO-TEK H27D), allowing fine control of the cavity length. The shear piezo actuator, and a pair of in-vacuum aspherical lenses to couple light into and out of the cavity, are all mounted onto a titanium disk, which is hanged from the vacuum viewport housing. A microscope objective mounted outside the customized low-aberration viewport provides numerical aperture of 0.65, and enables high resolution images of atoms and manipulations of individual atoms by optical tweezers \cite{Tian2022}.

To characterize the cavity properties optically, we stabilize the cavity length by a $1560~\mathrm{nm}$ laser, and measure the cavity transmission spectroscopy by scanning the frequency of a weak $780~\mathrm{nm}$ probe beam. The transmission spectrum is shown in Fig.~\ref{fig:Opticalmeasurement}(a). The spacing between adjacent $\mathrm{TEM}_{00}$ resonances gives a free spectral range of $\nu_{\mathrm{FSR}}=89.7~\mathrm{GHz}$, corresponding to a cavity length
$L=c/(2\nu_{\mathrm{FSR}})=1.67~\mathrm{mm}$.
For the same longitudinal order, the frequency spacings between the $\mathrm{TEM}_{01}$ and $\mathrm{TEM}_{00}$ modes and between the $\mathrm{TEM}_{10}$ and $\mathrm{TEM}_{00}$ modes are $\Delta\nu_{\mathrm{trans},y}=66.5~\mathrm{GHz}$ and $\Delta\nu_{\mathrm{trans},x}=66.0~\mathrm{GHz}$, respectively. The transverse-mode spacing is determined by the Gouy phase with
\begin{equation}
\Delta\nu_{\mathrm{trans},i}=\nu_{\mathrm{FSR}}\frac{\psi_{g i}}{\pi},
\qquad
\psi_{g i}=\arccos\left(1-\frac{L}{R_i}\right),
\end{equation}
where $i=x,y$ denotes the two orthogonal transversal axes. Equivalently, the cavity resonance frequencies are
\begin{equation}
\nu_{qmn}=\frac{c}{2L}
\left[
q+\frac{m+1/2}{\pi}\psi_{g x}
+\frac{n+1/2}{\pi}\psi_{g y}
\right],
\end{equation}
where $q$ is the longitudinal mode index, and $m,n$ are the transverse mode indices. From the measured transverse-mode spacings, we extract radii of curvature of $R_x=1.00~\mathrm{mm}$ and $R_y=0.98~\mathrm{mm}$, with an average radius of curvature $\bar R=0.99~\mathrm{mm}$. The corresponding cavity-mode waist at the cavity center is
$w_0=
\sqrt{
\frac{\lambda}{2\pi}
\sqrt{L(2\bar R-L)}
}
=9.5~\mu\mathrm{m}$.
The cavity linewidth is obtained from the ring-down measurement in Fig.~\ref{fig:Opticalmeasurement}(b), yielding caviy linewidth $\kappa=0.94\pm0.09~\mathrm{MHz}$. This gives a finesse
$\mathcal{F}=\frac{\nu_{\mathrm{FSR}}}{\kappa}=9.5\times10^4$.
Combining the measured finesse and mode waist, we obtain the optically-characterized single-atom cooperativity $\eta_{\mathrm{cav}}=24\mathcal{F}/[\pi(kw_0)^2]\simeq125$, where $k$ is the wave vector for 780~nm laser. This value is in good agreement with the cavity design and provides an independent optical benchmark for the cavity cooperativity.

We then load and position individual atoms inside the cavity mode through a transfer-and-rearrangement sequence. Neutral rubidium-87 atoms are first cooled in a magneto-optical trap (MOT) located approximately $3~\mathrm{mm}$ above the cavity center. After polarization-gradient cooling (PGC), the atoms are loaded into a horizontal optical dipole trap formed by a $1064~\mathrm{nm}$ laser. This dipole trap is vertically steered by an acousto-optic deflector (AOD), allowing the atoms to be adiabatically transported to the cavity center. A one-dimensional $850~\mathrm{nm}$ optical tweezer array, projected through an objective with a numerical aperture of 0.65, is then turned on to capture atoms from the dipole trap. With polarization-gradient cooling light, light-assisted collisions prepare none or single atom in each tweezers. We subsequently image the tweezer array and use real-time feedback to rearrange the occupied tweezers into the desired target configuration\cite{PhysRevApplied.19.054032,2gwz-65w1}. The rearranged atoms in optical tweezers are then transferred into an intracavity optical lattice formed by an $808~\mathrm{nm}$ laser. This lattice is frequency locked to the cavity and mode matched to the cavity field, thereby fixing the atoms at the antinode of the $780~\mathrm{nm}$ coupling. In the present setup, 21 individually controlled atoms can be rearranged into the cavity mode and coupled simultaneously to the maximally coupled position of the cavity field.

To characterize the atom-cavity coupling in the integrated platform, we measure the cavity transmission spectra for a controlled number of trapped atoms sitting at the maximally coupled position of the cavity. After atoms are rearranged and loaded into $808~\mathrm{nm}$ intracavity lattice at the antinode of $780~\mathrm{nm}$ coupling, a fluorescence image is taken to verify the prepared atom configuration. The atoms are then optically pumped into the stretched ground state of $5S_{1/2}\ket{F=2,m_F=2}$ in a bias magnetic field along the cavity at $5~\mathrm{Gauss}$ (see Appendix A for details). During the transmission spectroscopy measurement, the cavity resonance frequency $\omega_c$ is tuned near the Zeeman-shifted cycling transition $5S_{1/2}\ket{F=2,m_F=2}\rightarrow5P_{3/2}\ket{F'=3,m_{F'}=3}$, and the atom-cavity system is probed with a weak $\sigma^{+}$ polarized light of frequency $\omega$ that incidents onto the cavity. Then a second fluorescence image is taken to confirm that the atoms are all preserved. For each target atom number, the probe frequency is scanned across the cavity resonance, and the transmitted photons through the cavity are filtered by a $\sigma^+$ polarization filter and collected with a single-photon counting module (SPCM, Excelitas SPCM-AQRH-14-FC). Each data point is averaged over 20 experimental repetitions.

Representative transmission spectra for the empty cavity and for one, three, twelve, and sixteen trapped atoms are shown in Fig.~\ref{fig:spectra}. The empty-cavity spectrum in Fig.~\ref{fig:spectra}(b) exhibits a doublet structure with a splitting of $1.3~\mathrm{MHz}$. This splitting mainly originates from mirror-coating-induced birefringence, likely associated with residual stress in the dielectric coating. The contribution expected from the measured mirror ellipticity is only $\Delta \nu = \frac{\nu_{\mathrm{FSR}}}{\pi k}\frac{R_x-R_y}{R_xR_y}=72~\mathrm{kHz}$, much smaller than the observed splitting~\cite{Uphoff2015}. We therefore explicitly include this polarization splitting when calibrating the empty-cavity transmission and when converting the detected photon counts into normalized transmission.

With atoms loaded into the cavity mode, the spectra display the characteristic structure of the strongly coupled atom-cavity system. Fig.~\ref{fig:spectra}(c) shows the transmission spectrum for a single atom in cavity. Fitting the spectrum with the cavity-QED model gives an effective single-atom cooperativity of $\eta_{\mathrm{spec},N=1}=106\pm 3$, slightly below the optically-characterized value $\eta_{\mathrm{cav}}$. We attribute the difference between $\eta_{\mathrm{cav}}$ and $\eta_{\mathrm{spec}}$ to the spatial spreading of atoms due to finite atomic temperature. This spatial spreading leads to both lower atom-cavity coupling relative to that at the antinode of the cavity mode, and non-uniform AC stark shift of the ground state for atoms trapped in the lattice (see Appendix A for details). 

Figs.~\ref{fig:spectra}(d)--(f) show the transmission spectra for intracavity atom number $N=3$, 12, and 16. As the atom number increases, the Rabi splitting grows and the peak transmission is further suppressed, consistent with the collective cavity-QED response. The fitted effective single-atom cooperativity decreases modestly with atom number, from $\eta_{\mathrm{spec},N=1}=106\pm 3$ to $\eta_{\mathrm{spec},N=16}=91.4\pm 0.8$.
Several effects can contribute to this trend. First, atoms located off the center of the cavity mode experience a reduced coupling strength because of the Gaussian transverse profile of the cavity field. For the present tweezer spacing of $10.8~\mu\mathrm{m}$ and cavity Rayleigh range of $371~\mu\mathrm{m}$, the reduction in atom-photon coupling across a 16-atom array is estimated to be $1.8\%$ when the array is centered on the cavity mode, and about $4\%$ when the array center is displaced by $50~\mu\mathrm{m}$. Second, the transfer from the optical tweezers to the intracavity lattice may place some atoms one or two lattice sites away from the intended antinode. For the $808~\mathrm{nm}$ lattice, such axial placement errors reduce $\eta$ by $1.4\%$ and $5.4\%$ for one-site and two-site displacements, respectively. Third, imperfect optical pumping can leave a small fraction of atoms in undesired Zeeman sublevels. Under the $5~\mathrm{Gauss}$ bias field used for spectroscopy, these atoms are detuned from the cycling transition and couple more weakly because of the corresponding Clebsch--Gordan coefficients. As the atom number increases, the spectra peak from such imperfect optical pumping is no longer separated from the Rabi splitting peak, leading to a modest reduction of the extracted average single-atom cooperativity and a broadening of the observed spectra peak.

We therefore report $\eta_{\mathrm{cav}}$ as the optical figure of merit for the cavity, and $\eta_{\mathrm{spec}}$ as an in-situ trapped-atom verification of the atom-cavity strong coupling.
 Together with the ability to couple up to 200 individually controlled atoms to the cavity, these measurements demonstrate that the present architecture combines large number of atom array with ultra-high single-atom cavity cooperativity.

As a further demonstration of the cavity--atom interface, we measure the single-atom hyperfine-states of the $F=1$ and $F=2$ manifolds of the $5S_{1/2}$ ground state through the cavity transmission. In a $5~\mathrm{Gauss}$ bias magnetic field applied along the cavity axis, the cavity is tuned to the Zeeman-shifted cycling transition $5S_{1/2}\ket{F=2,m_F=2}\rightarrow5P_{3/2}\ket{F'=3,m_{F'}=3}$.
A single atom is trapped at an antinode of the intra-cavity lattice. After polarization-gradient cooling, optical-pumping light prepares the atom in the $F=2$ manifold. Alternatively, the $F=1$ state is prepared by turning off the repump light addressing the $5S_{1/2}\ket{F=1}\rightarrow5P_{3/2}\ket{F'=2}$ transition $10~\mathrm{ms}$ before the cooling light is turned off, allowing the population to accumulate in the $F=1$ manifold. We then inject a $100~\mathrm{\mu s}$ $\sigma^+$-polarized probe pulse through one cavity mirror at a probe--cavity detuning $\delta$ of $0.6~\mathrm{MHz}$. The transmitted light passes through a polarizing beam splitter (PBS), and we use two SPCMs to collect photons from both the transmitted and reflected ports of the PBS, summing their counts to obtain the total detected photon number. For state measurement, we bin the detected photons into $4~\mathrm{\mu s}$ time windows over the entire pulse duration and use the counts from these windows as the readout signal. After each probe pulse, we recool and optically pump the atom to prepare it for the next cycle. After 20 such cycles, a second fluorescence image is acquired to determine whether the atom remains trapped. The data-processing procedures and the parameters optimizations are discussed in Appendix C.

Figure~\ref{fig:fastreadout}(a) shows the single-atom cavity-transmission spectra for the two prepared hyperfine states. For $F=2$, resonant atom--cavity coupling produces vacuum Rabi splitting and strongly suppresses the transmission near the bare-cavity resonance. By contrast, an atom in $F=1$ is far detuned from the probe transition, and its transmission spectrum coincides with that of the empty cavity. We choose $\delta=0.6~\mathrm{MHz}$ as the probing frequency of the state measurement, where the cavity transmission ratio of the two hyperfine states are optimized. Figure~\ref{fig:fastreadout}(b) shows the histogram of the transmission photon-count in the logarithmic scale. The low-count distribution corresponds to atom in $F=2$, whereas the high-count distribution corresponds to atom in $F=1$. For a discrimination threshold $n_{\mathrm{th}}$, events with $n\leq n_{\mathrm{th}}$ are classified as $F=2$, whereas those with $n>n_{\mathrm{th}}$ are classified as $F=1$. Assuming equal prior probabilities for the two states, we define the state-measurement fidelity as
\begin{equation}
\mathcal{F}_{\mathrm{read}}=1-\frac{1}{2}\left[
\frac{N_{F=2}(n>n_{\mathrm{th}})}{N_{F=2}}
+\frac{N_{F=1}(n\leq n_{\mathrm{th}})}{N_{F=1}}
\right],
\end{equation}
where $N_{F=i}$ is the number of trials prepared in state $F=i$, and $N_{F=2}(n>n_{\mathrm{th}})$ and $N_{F=1}(n\leq n_{\mathrm{th}})$ are the corresponding numbers of incorrect readout outcomes. Applying this definition directly to the measured distributions, we find that $n_{\mathrm{th}}=2$ minimizes the observed readout error and yields a state-measurement fidelity of $\mathcal{F}_{\mathrm{read}}=99.925^{+15}_{-18}\%$ for the $4~\mathrm{\mu s}$ counting window. The second fluorescence image quantifies the atom survival rate in each counting window to be $99.996\%$, thus atom loss during the readout is negligible.

%\textcolor{blue}{Exploiting the strong atom--photon coupling, we realize a 4~$\mathrm{\mu s}$ state detection of an atom prepared in either $5S_{1/2}\ket{F=2}$ or $5S_{1/2}\ket{F=1}$ in the cavity. Specifically, the cavity frequency is first locked to the $5S_{1/2}\ket{F=2,m_F=2}\rightarrow5P_{3/2}\ket{F'=3,m_{F'}=3}$ transition under a 5 Gauss magnetic field applied along the cavity axis. Following the procedure described in the previous section, a single atom is captured at the antinode of the intracavity lattice and optically pumped into the $5S_{1/2}\ket{F=2}$ state after PGC cooling, while the $5S_{1/2}\ket{F=1}$ state is prepared by switching off the repump light driving the $5S_{1/2}\ket{F=1}\rightarrow5P_{3/2}\ket{F'=2}$ transition 10~ms before the cooling light is turned off. The prepared atom is then interrogated by a 10~$\mu$s $\sigma^+$ probe pulse with detuning $\delta=0.6$~MHz is injected through one of the cavity mirrors, and the transmitted photons are collected by an SPCM. Upon completion of the probe pulse, the atom survival probability is subsequently evaluated using the second fluorescence image introduced in the previous section.}

\begin{figure}
    \centering
    \includegraphics[width=0.95\linewidth]{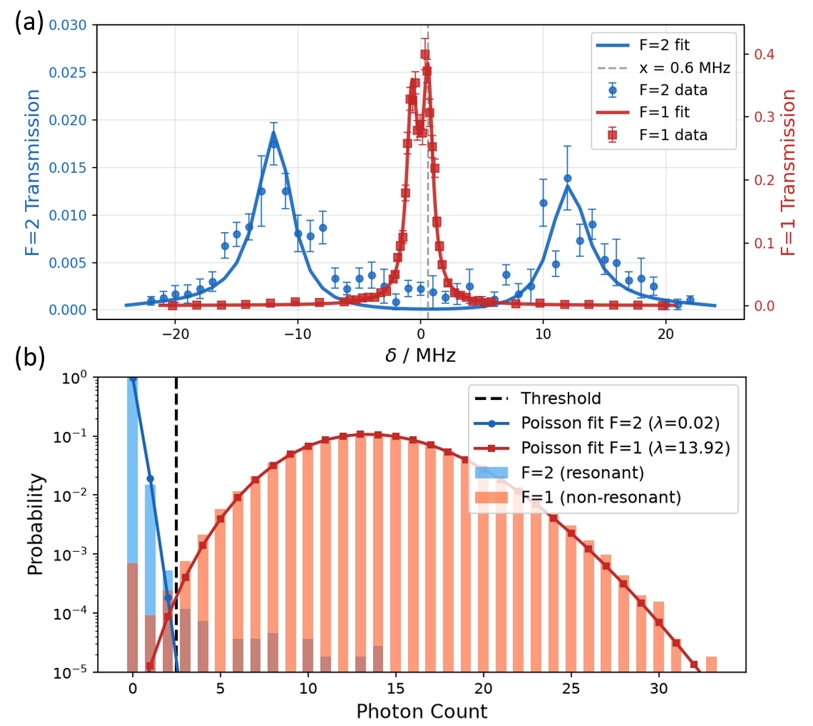}
        \caption{Fast cavity-based hyperfine-state readout of a single atom. (a) Cavity transmission as a function of the probe-cavity detuning $\delta$ for an atom prepared in $F=2$ (blue dots, left axis) or $F=1$ (red square, right axis). Markers show the measured transmission and solid curves are fits to the corresponding spectra. Resonant coupling of the $F=2$ atom produces a vacuum-Rabi doublet, whereas the far-detuned $F=1$ atom gives an empty-cavity-like central resonance. The vertical dashed line marks the readout probe-cavity detuning $\delta=0.6~\mathrm{MHz}$. (b) Histogram of cavity transmission photon counts detected with $4~\mathrm{\mu s}$ probe time. The blue and red histograms correspond to atoms prepared in $F=2$ and $F=1$, respectively; solid curves are Poisson fits with mean counts $\bar n_2=0.02$ and $\bar n_1=13.92$. The vertical dashed line denotes the optimum threshold $n_{\mathrm{th}}=2$: outcomes with $n\leq2$ are assigned to $F=2$, and those with $n>2$ to $F=1$. Applying this threshold to the measured distributions gives a state-measurement fidelity of $99.925^{+15}_{-18}\%$.}
%    \caption{Fast cavity-based atomic-state readout. (a) Cavity transmission spectra for a single atom in the $F=2$ (blue) and $F=1$ (red) states. The $F=2$ spectrum shows vacuum Rabi splitting, while the $F=1$ spectrum nearly overlaps with the empty-cavity transmission. The readout operating point is chosen at a probe-cavity detuning of $\delta=0.6~\mathrm{MHz}$ (dashed line). (b) Probability distribution of detected photon counts within $4~\mathrm{\mu s}$ of probe time. Red/blue points correspond to the $F=1/F=2$ states. Poisson fits yield mean photon numbers of 13.92 and 0.02, respectively, showing clear separation between the two distributions. The dashed line indicates the threshold chosen for state discrimination.}
    \label{fig:fastreadout}
\end{figure}

%\textcolor{blue}{Figs~\ref{} shows the corresponding detection results. The left peak corresponds to atoms in the $5S_{1/2}\ket{F=2}$ state, while the right peak corresponds to atoms in the $5S_{1/2}\ket{F=1}$ state. The detection infidelity is calculated as
%\begin{equation}
%\mathcal{I}=\frac{N_{1,>thres}+N_{0,<thres}}{N_{total}},
%\end{equation}
%where $N_{1,>thres}$ denotes the number of measurements with collected photon counts exceeding the threshold for atoms in the $5S_{1/2}\ket{F=2}$ state, and $N_{0,<thres}$ denotes the number of measurements with photon counts below the threshold for atoms in the $5S_{1/2}\ket{F=1}$ state. By optimizing the threshold, the maximum detection fidelity reaches 99.93\% at a threshold of two counts. Here, the atom loss probability during detection is negligible, as determined by postselection.}

In conclusion, we have realized a cavity-integrated neutral-atom array that combines individual-atom control with strong atom--photon coupling in a millimeter-scale, side optically accessible cavity geometry. Optical characterization gives the cooperativity of $\eta_{\mathrm{cav}}=125\pm13$, while trapped-atom spectra show the cooperativity $\eta_{\mathrm{spec}}=106\pm3$, confirming that high cooperativity is retained under actual array-operation conditions; up to 21 individually positioned atoms are coupled to the cavity antinode simultaneously. The resulting state-dependent transmission contrast enables single-atom hyperfine-state readout with a measured fidelity of $99.925\%$ in a $4~\mathrm{\mu s}$ photon-counting window, and the measured atom loss rate is $0.004\%$. By combining high cooperativity, multi-atom scalability, and fast high-fidelity readout while preserving high-NA optical access, this platform connects local tweezer control with the global connectivity of cavity QED and provides a route toward cavity-assisted measurement, photon-mediated interactions, and collective entanglement generation in programmable atom arrays~\cite{wu_engineering_2026,ygdy-hrgj,ramette_carving_2025,yu_efficient_2026,Zhang2024Quantum}.

%In conclusion, we have realized a cavity-integrated atomic-array platform that combines millimeter-scale geometry, flexible optical access, and a optically-characterized single-atom cooperativity of $\eta_{\mathrm{cav}}=125\pm13$. Trapped-atom transmission spectra yield an effective cooperativity of $\eta_{\mathrm{spec}}=106.4\pm 3.0$, providing an in-situ verification of strong coupling under actual atom array operating conditions.
%The key element to enable such platform is a two-step mirror-fabrication method that achieves sub-millimeter radii of curvature together with ultra-low surface roughness, thereby enables ultra-high-cooperativity cavity in a geometry compatible with tweezer-based individual atom array manipulations. The resulting platform provides a new operating regime for cavity QED and a versatile route toward cavity-mediated many-body quantum control and mea-
%surement. This enables investigations for novel quantum many-body dynamics\cite{wu_engineering_2026,ygdy-hrgj}, fast and high-fidelity quantum state readout, and efficient preparation of highly entangled states for quantum metrology and quantum information
%processing \cite{ramette_carving_2025,yu_efficient_2026,Zhang2024Quantum}.

We thank Tao Zhang and Tian Tian for experimental guidance of cavity mirror substrate fabrication. This work is supported by National Key Research and Development Program of China (2023YFA1406702, 2021YFA1400904, 2021YFA0718303), 
National Natural Science Foundation of China (92576208, 92476110) and Tsinghua University Initiative Scientific Research Program and Dushi Program.

\bibliographystyle{apsrev4-1}
\bibliography{references_abbrev}

\appendix

\section{Atom-cavity Transmission Spectra}

We measure the atom-cavity transmission spectra for a controlled number of trapped atoms, to characterize the atom-cavity coupling in the integrated platform. Atoms are rearranged and loaded into $808~\mathrm{nm}$ intracavity lattice at the cavity antinode for $780~\mathrm{nm}$ coupling. A fluorescence image is taken through microscope objective to verify the prepared atom configuration. 

The atoms are then optically pumped into the stretched ground state in a bias magnetic field transverse to the cavity axis. The pumping sequence uses $\sigma^+$ light on the $\ket{F=2}\rightarrow\ket{F'=2}$ transition, together with depumping light addressing the $\ket{F=1}\rightarrow\ket{F'=2}$ transition. The bias magnetic field is subsequently rotated adiabatically to the cavity axis and set to $5~\mathrm{Gauss}$. 

During transmission spectra measurement, the cavity resonance frequency $\omega_c$ is tuned near the Zeeman-shifted cycling transition $5S_{1/2}\ket{F=2,m_F=2}\rightarrow5P_{3/2}\ket{F'=3,m_{F'}=3}$, and the atom-cavity system is probed with weak right-circularly polarized light of frequency $\omega$ that incident onto the cavity. 

For each target atom number, the probe frequency is scanned across the cavity resonance. And we use a quarter-wave plate and a polarizing beam splitter (PBS) to filter out the residual $\sigma^-$ polarized light component in the photons transmitted through the cavity (see Fig.~\ref{fig:spectra}(a)). This polarization filter at the output of the cavity affects the transmission peaks of the polarization splitting observed in Fig.~\ref{fig:spectra}(b) .

In the transmission spectra measurement sequence, each probe duration takes $30~\mathrm{\mu s}$, and the incident light intensity in all cases remains the same, with the average photon number of 3 in the empty cavity.

The efficiency of the photon detection path at the output of the cavity is calibrated as below. Each cavity mirror has $19~\mathrm{ppm}$ transmission and $13~\mathrm{ppm}$ loss for incident light, and introduce a transmission efficiency of intra-cavity photon at the output interface of the cavity to be $59.3 \%$. The dichroic mirror, fiber coupling and bandpass filters in the detection path introduce a total transmission of $81 \%$. And the SPCM has an efficiency of $65 \%$. As a result, a total detection efficiency of $31.2 \%$ is achieved.

To analyze the atom-cavity transmission spectra, we fit the spectra using three free parameters: the atom-cavity detuning $\Delta$, the light-cavity detuning $\delta$ and the single-atom cooperativity $\eta$.

The thermal spatial distributions of the atoms are crucial in the fitting, because the local atom-cavity coupling strength $g$, and the effective detuning $\Delta$ considering the AC stark shift, both depend on the spatial spreading of the atoms in the cavity mode and the trapping lattice. 

For a given value of cavity cooperativity $\eta$, the transmission spectrum of N atoms in the cavity is described by \cite{TANJISUZUKI2011201}
\begin{equation}
    \hat{T}(\delta,\Delta)
    =
    \left|
    \frac{1}
    {
    1-2i\delta/\kappa
    +
    \dfrac{N\eta}{1+2i(\Delta-\delta)/\Gamma}
    }
    \right|^2.
\end{equation}

When the atom is displaced away from the cavity axis by a distance r and from the $780~\mathrm{nm}$ antinode by a distance y, the single-atom cooperativity is given by $\eta(r,y)=\eta_{0}\cdot \mathrm{exp}(-2r^2/w^2_0)\mathrm{cos}^2(ky)$, and the atom–cavity detuning is $\Delta(r,y) =\Delta -\Delta_{\mathrm{AC}}[1-\mathrm{exp}(-2r^2/w^2_0)\mathrm{cos}^2(k_ly)] $, where $\eta_{0}$ and $\Delta $ are the single‑atom cooperativity and detuning experienced by an atom on the cavity axis and at the antinode of the $780~\mathrm{nm}$ cavity mode, $\Delta_{\mathrm{AC}}$ is the AC Stark shift at the bottom of the $808~\mathrm{nm}$ optical lattice, and $k (k_{l})$ are the $780~\mathrm{nm}$ ($808~\mathrm{nm}$) wavenumbers. If the atomic temperature is T, with the trap frequencies of the atom in the optical lattice in the axial and radial direction being $\omega_{l}$ and $\omega_r$, the atomic distribution in the axial direction is given by $P(y)=\frac{1}{Z}\sum_n \mathrm{exp}(-n\hbar \omega_l/k_B T)|\Psi_{n}(y)|^2$, and the atomic distribution in the radial direction is $P'(r)=\frac{1}{Z'}\sum_n \mathrm{exp}(-n\hbar \omega_r/k_B T)|\Psi'_{n}(r)|^2$, where $\Psi_{n}(y)$ and $\Psi'_{n}(r)$ are the wave functions of the n-th excited state in each direction, respectively. Z and Z' are the corresponding partition functions. Considering the universal temperature, the fitting expression becomes
\begin{widetext}
\begin{equation}
    \hat{T}(\delta, \Delta)=\int\int P(y)P'(r)
    \left|
    \frac{1}
    {
    1-2i\delta/\kappa
    +
    \dfrac{N\eta(r,y)}{1+2i(\Delta(r,y)-\delta)/\Gamma}
    }
    \right|^2 2\pi rdrdy.
    \label{eqn:etafitting}
\end{equation} 
\end{widetext}

Measuring the atom temperature by release and recapture thermometry in lattice, we find that the atoms are at $14~\mu$K before transmission spectra probing, and between $13~\mu$K and $50~\mu$K after transmission spectra probing. The cavity cooling and cavity heating processes during probing leads to different atomic temperatures after transmission spectrum probing at different detuning $\delta$. In the fittings in Fig.~\ref{fig:spectra}, we assume a universal atomic temperature of $13~\mu$K for both radial and axial direction, as $13~\mu$K is the lower bound of the atomic temperature. Assuming atoms are at this lower bound universal temperature, we fit transmission spectrum data with the fitting expression in equation \ref{eqn:etafitting} and gives the lower bound of the spectral cavity cooperativity $\eta_{\mathrm{spec}}$ with which the atoms actually interact.

In particular, in the empty cavity, the polarization-split cavity eigenmodes are linearly polarized \cite{Uphoff2015}. As a result, when the transmitted light passes through the polarization-analysis optics and is detected by the single-photon counting module (SPCM), the measured photon number is given by
\begin{equation}
    n_{\mathrm{ph}}
    =
    \frac{n_p}{4}
    \left|
    \frac{1}{1-2i(\delta+\delta_p/2)/\kappa}
    +
    \frac{1}{1-2i(\delta-\delta_p/2)/\kappa}
    \right|^2 ,
\end{equation}
where $n_p$ is the incident pump photon number, and $\delta_p$ is the polarization-mode splitting. This expression is used to fit the empty-cavity transmission spectrum and calibrate the probe intensity, thereby converting the measured photon counts into a normalized transmission spectrum.

\section{Atom Thermometry in Lattice by Release and Recapture}

To measure the temperature of the atoms in the lattice, we used the release and recapture method. The trapping lattice is turned off, the atoms evolve freely with gravity for a variable release time, and the lattice is turned on again to recaptured atoms. By comparing the experimental probability of atomic recapture with that simulated by the Monte Carlo approach, the radial temperature of the atoms is extracted \cite{He2011CPB,Fuhrmanek2010NJP,Kaufman2012PRX}.

\begin{figure*}[t]
    \centering
    \includegraphics[width=0.95\linewidth]{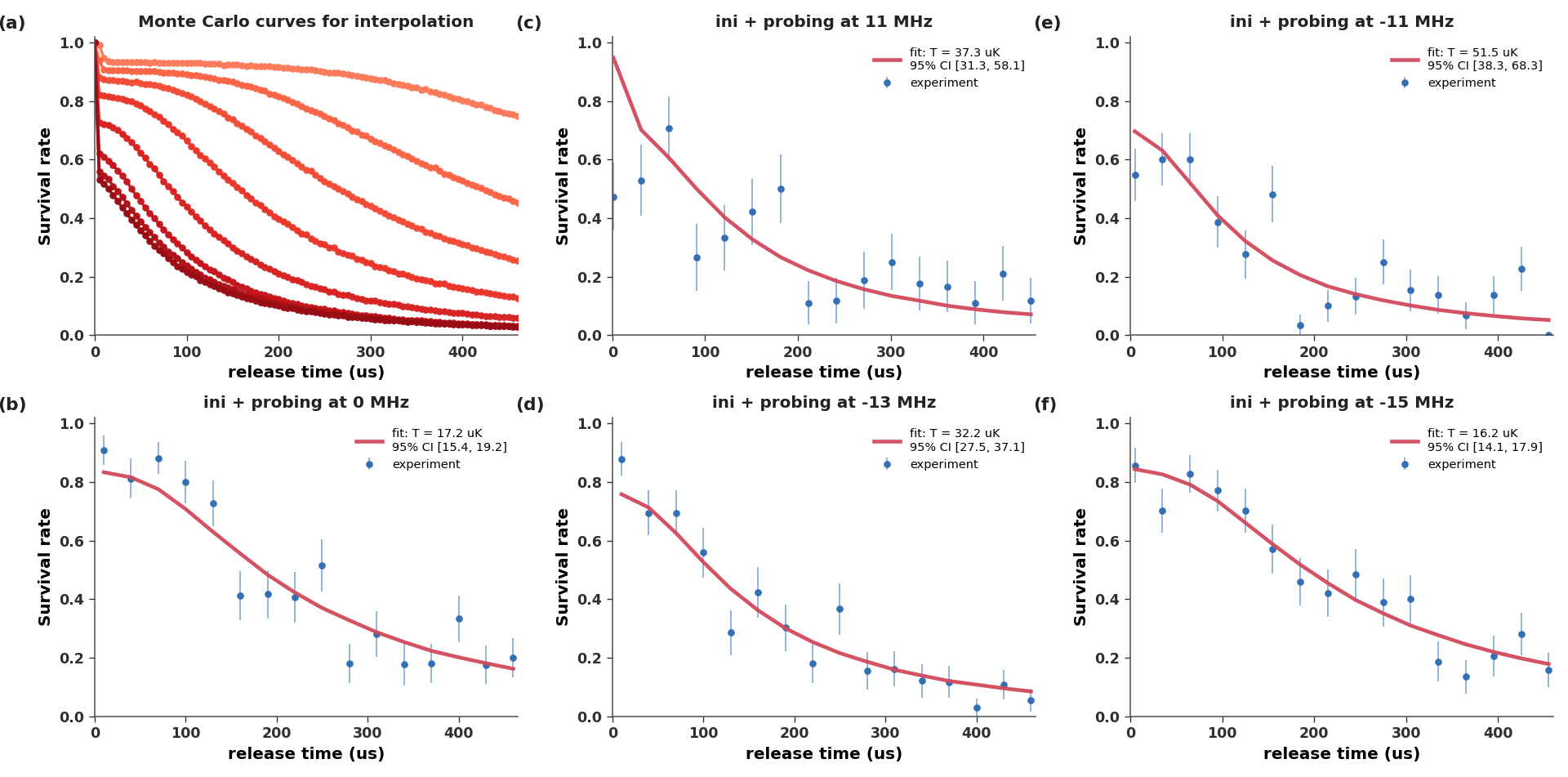}
    \caption{Lattice release-and-recapture thermometry. 
(a) Simulated recapture probability as a function of release time for different temperatures, obtained from Monte Carlo calculations and used for interpolation. (b)--(f) Experimental recapture data under different 
% \st{preparation and }
probing conditions, together with interpolation-based fits used to extract the atomic temperature and its confidence interval. 
% \st{Here ``pol'' denotes optical pumping and magnetic-field rotation used to prepare the atoms in the desired spin state. ``Probing'' denotes application of the cavity probe field with the same parameters as those used in the single-atom transmission-spectrum measurement. }
Error bars represent statistical uncertainties of the measured recapture probabilities.
In this series of measurement, two resonant peaks locate at -13 MHz and 11 MHz.}
    \label{fig:recapture}
\end{figure*}

In Monte Carlo simulations, for each release time, we use $10^4$ trajectories and obtain the recapture probability as a function of the release time for a given temperature. By interpolation over a range of temperatures, we generate a family of recapture curves that can be used to fit the experimental data and extract the temperature (Fig.~\ref{fig:recapture} (a)).

The measurements are performed after atomic state preparation, and after cavity probing at different probe beam detunings. In principle, cavity-induced cooling or heating would lead to different atomic temperature at different probe detuning, so the temperature should vary across the data points in transmission spectra. Thus, we measure the atomic temperature after transmission spectra probe at five typical detunings: on resonance $\delta=\Delta=0$ (Figs.~\ref{fig:recapture} b), at the peak of both Rabi splitting peaks $\Delta=0$ and $\delta=11~\mathrm{MHz}$ or $\delta=-13~\mathrm{MHz}$ (Figs.~\ref{fig:recapture} c,d), and on the cooling and heating slope of the blue detuned Rabi splitting peak (Figs.~\ref{fig:recapture} e,f).

\begin{table*}[htbp]
\centering
\caption{Measured radial temperatures of atoms for different probing detunings and intracavity atom numbers under two conditions: before and after thermalization between radial and axial temperature. Dashes (--) indicate no available data.}
\label{tab:temps}
\begin{tabular}{lcc}
\toprule
Feature & Before thermalization ($\mu$K) & After thermalization ($\mu$K) \\
\midrule
\multicolumn{3}{c}{Single atom} \\
Initialization & 13.6 & -- \\
Probe on resonance ($\delta = 0$~MHz) & 17.2 & 18.2 \\
Probe on positive peak ($\delta = 11$~MHz) & 37.3 & -- \\
Probe on negative peak left slope ($\delta = -15$~MHz) & 16.2 & 19.4 \\
Probe on negative peak ($\delta = -13$~MHz) & 32.2 & -- \\
Probe on negative peak right slope ($\delta = -11$~MHz) & 51.5 & 35.5 \\
\midrule
\multicolumn{3}{c}{Six atoms} \\
Probe on negative peak left slope($\delta = -31$~MHz) & 12.7 & -- \\
Probe on negative peak ($\delta = -29.5$~MHz) & 12.5 & -- \\
Probe on negative peak right slope ($\delta = -28$~MHz) & 15.5 & -- \\
Probe on positive peak left slope ($\delta = 28$~MHz) & 14.7 & 14.7 \\
Probe on positive peak ($\delta = 29.5$~MHz) & 12.9 & 17.7 \\
Probe on positive peak right slope ($\delta = 31$~MHz) & 15.9 & 17.5 \\
\bottomrule
\end{tabular} \label{tab:temperature}
\end{table*}

The release and recapture temperature measurement are effectively measuring radial temperature of atoms in lattice. To estimate the axial temperature, we do the measurement after waiting for 50~ms following the transmission spectra probing, when the axial and radial motion are completely thermalized, as shown in Tab.~\ref{tab:temperature}. Due to the complex temperature dependence on probe detunings, we simplify the temperature dependence model and reasonably assume that all atoms are at $13~\mu$K for the fitting in the transmission spectra to give a lower bound of the atom observed cavity cooperativity $\eta_{\mathrm{spec}}$.

\section{Atomic-State Measurement Using a Cavity}

In this appendix, we present the parameter optimizations and detailed analysis of the cavity-based state-readout measurements. We elaborate the parameter optimizations for state measurement, state flipping determination in the data analysis, and how the readout fidelity is analyzed.

\subsection*{Parameter Optimizations for State Measurement}

To optimize the experimental parameters to get high fidelity and fast atomic state measurement, we analyze the cavity probe light induced saturation and state redistributions.

The dependence on probe power can be made explicit by writing the mean detected count as
\begin{equation}
    \mu_i(I,\tau)=\tau\,[R_i(I)+R_{\mathrm{bg}}],
    \label{eq:mean_count_rate}
\end{equation}
where $I$ is the incident probe intensity, $R_i(I)$ is the state-dependent transmitted-photon detection rate, and $R_{\mathrm{bg}}$ accounts for detector dark counts and other background. In the weak-probe regime, the mean intracavity photon number is much smaller than unity, atomic saturation is negligible, and the response is linear: $R_i(I)\simeq\alpha_i I$. If $R_{\mathrm{bg}}\ll R_i$ and $\tau$ is adjusted so that $\mu_1$ remains fixed while $I$ is varied, then $I\tau$ remains fixed and
\begin{equation}
    \mu_2\simeq \frac{\alpha_2}{\alpha_1}\mu_1.
    \label{eq:linear_scaling}
\end{equation}
Thus, provided that the full count distributions remain Poissonian and no atomic state flipping occurs during the probe, the shot-noise-limited discrimination fidelity is approximately independent of the separate choices of $I$ and $\tau$ at fixed $\mu_1$.

\begin{figure*}[t]
    \centering
    \includegraphics[width=0.95\linewidth]{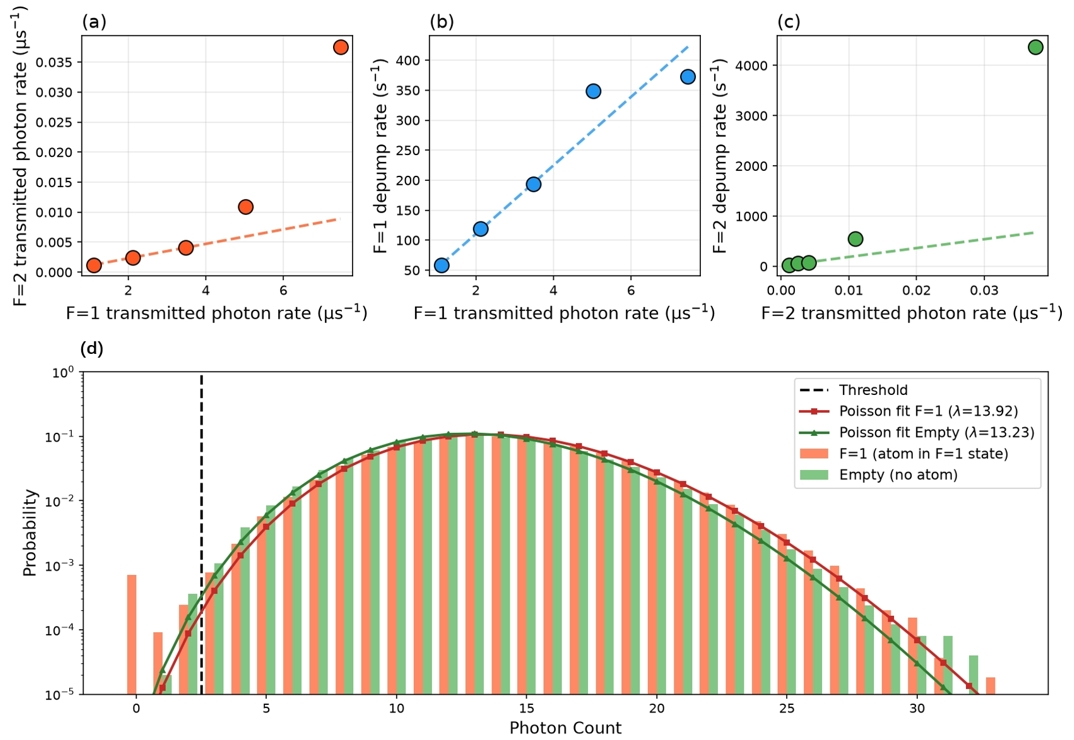}
    \caption{Probe-power dependence and error mechanisms of cavity-enhanced atomic-state readout. The cavity is resonant with the $F=2\rightarrow F'=3$ transition, and the probe--cavity detuning is fixed at $\delta=0.6~\mathrm{MHz}$. (a) SPCM detected cavity transmission photon rate for an atom prepared in $F=2$ plotted against that for an atom prepared in $F=1$. Each point corresponds to a different combination of probe power and integration time, with the mean integrated $F=1$ photon count held fixed. The dashed line indicates the weak-probe linear-response trend; the highest-power point departs strongly from this scaling due to saturation and optical depumping. (b) Measured $F=1\rightarrow F=2$ depumping rate versus the $F=1$ cavity transmission photon count rate. The rate is approximately linear at all count rates (dashed line). (c) Measured $F=2\rightarrow F=1$ depumping rate versus the $F=2$ cavity transmission photon count rate. The sharp rise of the highest-power point above the linear extrapolation shows strongly nonlinear probe-induced state redistribution. (d) Photon-count distributions during a $4~\mathrm{\mu s}$ integration window for an atom prepared in $F=1$ (red) and for the empty cavity (green), measured under the same probe conditions. The corresponding Poisson fits have mean counts $\bar n_1=13.92$ and $\bar n_{\mathrm{empty}}=13.23$. The vertical dashed line marks $n_{\mathrm{th}}=2$; the excess probability of the $F=1$ data in the low-count bins contributes directly to the measured readout error.}
%    \caption{Dependence of cavity-enhanced atomic-state readout performance on probe intensity. (a) Correlation between SPCM photon count rates for the  F=1 and F=2  states under varying probe intensities, with the cavity resonant with the $F=2\rightarrow F'=3$ transition and the probe-cavity detuning fixed at $\delta=0.6~\mathrm{MHz}$. (b)Depumping rate from F=1 to F=2 as a function of probe intensity; the x-axis is the photon count rate for the corresponding F=1 state. (c) Depumping rate from F=2 to F=1 as a function of probe intensity; the x-axis is the photon count rate for the corresponding F=2 state. (d) Probability distributions of transmitted photon counts within $4~\mathrm{\mu s}$ for the F=1 state (red) and the empty-cavity reference (green) under the same probe intensity.}
    %The fits yield a single-atom cooperativity of $\eta = 110\pm10$.}
    \label{fig:depump}
\end{figure*}

To test the limits of this scaling, we vary the probe intensity and integration time while holding the measured $F=1$ mean count fixed, and record the corresponding $F=2$ mean count. In the weak-probe regime [linear region in Fig.~\ref{fig:depump}(a)], $\bar n_2$ remains nearly constant, as expected from Eq.~\eqref{eq:linear_scaling}. The operating point used in the main text corresponds to the middle data point in Fig.~\ref{fig:depump}(a).  At high probe intensity, $\bar n_2$ increases even though $\bar n_1$ is fixed. This departure from linear scaling is consistent with saturation of the resonantly coupled $F=2$ transition, which weakens the atom-induced suppression of cavity transmission. Probe-induced spontaneous scattering and optical pumping can additionally redistribute population among hyperfine or Zeeman states during the measurement. Both effects can alter not only the mean counts but also the tails of the distributions, and hence increase the errors defined in Eqs.~\eqref{eq:error_F1} and \eqref{eq:error_F2}.

To characterize this measurement-induced state flipping directly, we measure the depumping rates in both directions as functions of the corresponding transmitted-photon count rates. For atoms initially prepared in $F=1$, the $F=1\rightarrow F=2$ depumping rate increases approximately linearly at low and intermediate $F=1$ count rates [Fig.~\ref{fig:depump}(b)]. For atoms initially prepared in $F=2$, the $F=2\rightarrow F=1$ depumping rate remains comparatively small at low count rates and then rises sharply in the high-intensity regime, well above the linear extrapolation [Fig.~\ref{fig:depump}(c)]. This strong nonlinear increase occurs in the same high-power regime in which the $F=2$ transmitted-photon rate departs from the weak-probe scaling in Fig.~\ref{fig:depump}(a). The combined observations show that increasing the probe power not only reduces the state-dependent transmission contrast through saturation but also enhances probe-induced depump of the atomic state. The present measurements do not by themselves distinguish among the microscopic optical-pumping pathways, so we refer to these processes collectively as probe-induced state redistribution. The probe intensity and integration time must therefore be optimized jointly to balance photon shot noise against saturation and depumping.

As a result, we optimize the power of the cavity probe beam so that it sits at the third point in Fig.~\ref{fig:depump}(a)-(c), to guarantee that the cavity probe power is optimized for fast state measurement with minimum state flipping induced by state measurement.

We also compare the distribution of transmitted photon counts for atom measured in $F=1$, with the empty-cavity distribution obtained under otherwise identical probe conditions, and show the results in Fig.~\ref{fig:depump}(d). Their fitted mean counts are similar, $\bar n_1=13.92$ and $\bar n_{\mathrm{empty}}=13.23$. The histogram of transmitted photon counts for atoms in F=1 and empty cavity are similar, the extra small components in the low-count bins $n=0,1,$ and $2$ for $F=1$ case are due to state flipping in the measurement.

\subsection*{Determination of State Flipping in the State Measurement}

To increase the data requisition rate, each cavity transmission probe exposure consists of a 100 $~\mu$s probe pulse, with detected photons binned into 4~$\mu$s time windows. As this repetitive state measurement sequence increases the probability of state flipping in the measurement, we get rid of the extra state flipping case in the measurement data in data analyze by determining the time a state flipping happens.

To evaluate the readout fidelity and characterize probe-induced state changes, we first identify when, if at all, the atomic state flips between the two hyperfine states during the probe pulse. Once a flip occurs, the subsequent time windows no longer represent the initially prepared state; retaining them would introduce a large depumping-related error into the fidelity evaluation. To reliably identify the flip time and discard all bins after it, we apply a single-threshold binarisation followed by a three-consecutive-wrong-bin rule to the photon-count trajectories.

First, a prior threshold $C$ was determined from the crossing point of the $F=1$ and $F=2$ histograms. This threshold should agree with the final fidelity judging threshold. A bin was labelled ``high'' if its photon number exceeded $C$, and ``low'' otherwise. Under this classification, $F=1$ is expected to remain high throughout, and $F=2$ low.

A state flip is declared only when three consecutive bins are all classified into the wrong state relative to the initially prepared state; the flip time is assigned to the first bin of this three-bin window. We have verified that choosing two consecutive bins, or more than three, as the flip criterion does not change the extracted fidelity or depumping rates beyond the error bar. Once a flip is identified, we retain the first bin of the three-bin window and discard all subsequent bins. If the three consecutive wrong bins begin from the very first bin of the exposure, the exposure is classified as a preparation failure and discarded entirely. The preparation failure fraction is about $3\%$. Recording the time of the first state flip yields a depumping curve, from which we extract the preparation success probability and the depumping rates of the two atomic states.

The three-consecutive-wrong-bin rule is appropriate here because the depumping rate is low and the typical photon numbers of the two states are well separated from the threshold. A single-bin fluctuation is unlikely to persist for three consecutive bins, so this criterion effectively rejects isolated outliers while reliably identifying genuine flips.

\subsection*{Readout Fidelity and Poisson Limit}

Using the retained bins for $F=1$ and $F=2$, we define the empirical conditional distributions and their means as follows. Let $i\in\{1,2\}$ label the prepared hyperfine manifold $F=i$, let $n$ denote the number of photons detected within an integration time $\tau$, and let $N_i(n)$ be the number of experimental trials that yield $n$ counts after preparation in $F=i$. If $N_i=\sum_{n=0}^{\infty}N_i(n)$ is the total number of trials for that preparation, the empirical conditional distribution and its mean are
\begin{equation}
    \widehat{p}_i(n)\equiv \widehat{P}(n\mid F=i)
    =\frac{N_i(n)}{N_i},
    \qquad
    \bar n_i=\sum_{n=0}^{\infty}n\,\widehat{p}_i(n).
    \label{eq:readout_distribution}
\end{equation}
At the operating point indicated in Fig.~\ref{fig:fastreadout}(a), and for $\tau=4~\mu\mathrm{s}$, the measured distributions in Fig.~\ref{fig:fastreadout}(b) give $\bar n_1=13.92$ and $\bar n_2=0.02$. The large contrast arises because the $F=2$ state is resonantly coupled to the cavity and suppresses its transmission, whereas the $F=1$ state is far detuned from the probe transition and produces a transmission close to that of the empty cavity.

For a threshold $n_{\mathrm{th}}$, we assign an outcome to $F=2$ when $n\leq n_{\mathrm{th}}$ and to $F=1$ when $n>n_{\mathrm{th}}$. The two empirical conditional error probabilities are therefore
\begin{align}
    \widehat{\epsilon}_1(n_{\mathrm{th}})
    &=\widehat{P}(n\leq n_{\mathrm{th}}\mid F=1)
      =\sum_{n=0}^{n_{\mathrm{th}}}\widehat{p}_1(n),
      \label{eq:error_F1}\\
    \widehat{\epsilon}_2(n_{\mathrm{th}})
    &=\widehat{P}(n>n_{\mathrm{th}}\mid F=2)
      =\sum_{n=n_{\mathrm{th}}+1}^{\infty}\widehat{p}_2(n).
      \label{eq:error_F2}
\end{align}
More generally, for prior probabilities $\pi_1$ and $\pi_2$, with $\pi_1+\pi_2=1$, the average state-measurement fidelity is
\begin{equation}
    \widehat{\mathcal{F}}_{\mathrm{read}}
    =1-\pi_1\widehat{\epsilon}_1-\pi_2\widehat{\epsilon}_2.
    \label{eq:general_readout_fidelity}
\end{equation}
Throughout this work we use equal priors, $\pi_1=\pi_2=1/2$, and choose the threshold that maximizes Eq.~\eqref{eq:general_readout_fidelity}. The optimum $n_{\mathrm{th}}=2$ yields $\widehat{\mathcal{F}}_{\mathrm{read}}=99.925\%$. This value is evaluated from the measured histograms and therefore includes imperfect state preparation and any non-Poissonian tails present in the data; no correction for either contribution is applied. Atom loss during the $4~\mu\mathrm{s}$ readout window is below $0.004\%$ and does not measurably affect the quoted fidelity. For finite data sets, the statistical uncertainty can be obtained from binomial confidence intervals for the two observed error counts, $N_1\widehat{\epsilon}_1$ and $N_2\widehat{\epsilon}_2$, and propagated through Eq.~\eqref{eq:general_readout_fidelity}.

For comparison, if photon detection were a stationary Poisson process, the conditional count distribution would be
\begin{equation}
    p_i^{(\mathrm{P})}(n)=e^{-\mu_i}\frac{\mu_i^n}{n!},
    \qquad \mu_i=\mathbb{E}[n\mid F=i].
    \label{eq:poisson_distribution}
\end{equation}
The corresponding threshold errors are
\begin{align}
    \epsilon_1^{(\mathrm{P})}
    &=e^{-\mu_1}\sum_{n=0}^{n_{\mathrm{th}}}\frac{\mu_1^n}{n!},
    \label{eq:poisson_error_F1}\\
    \epsilon_2^{(\mathrm{P})}
    &=1-e^{-\mu_2}\sum_{n=0}^{n_{\mathrm{th}}}\frac{\mu_2^n}{n!}.
    \label{eq:poisson_error_F2}
\end{align}
Using $\mu_1=13.92$, $\mu_2=0.02$, and $n_{\mathrm{th}}=2$ gives $\epsilon_1^{(\mathrm{P})}=1.07\times10^{-4}$ and $\epsilon_2^{(\mathrm{P})}=1.31\times10^{-6}$, corresponding to an ideal photon-shot-noise-limited fidelity of $99.9946\%$. The lower measured value of $99.925\%$ is thus determined by the tails of the empirical distributions rather than by the Poisson fits to their mean counts.

% \textcolor{blue}{We also measured the transmitted photon number distributions for two reference cases: the atom prepared in the $F=1$ state and the empty cavity. As shown in Fig\ref{fig:depump}.(d), the two distributions exhibited comparable average photon numbers and similar overall shapes.}

% \textcolor{blue}{A closer inspection of the detailed distributions, however, reveals that—compared to the empty-cavity case—the $F=1$ case shows a notable increase in occurrence counts at low photon-number channels (0, 1, and 2). This discrepancy arises primarily from two factors. First, the initial state preparation is imperfect: a small fraction of atoms are not completely pumped into the $F=1$ manifold, leaving residual population in the $F=2$ state, which contributes additional photon scattering events during the measurement. Second, even for atoms initially prepared in $F=1$, off-resonant scattering or optical pumping induced by the probe beam can transfer some population into the $F=2$ manifold, thereby contributing to the low-photon-count events. These effects collectively produce a low-count tail in the $F=1$ distribution relative to the empty-cavity baseline, and constitute one of the dominant sources of measurement error in the current system. Improving the state preparation fidelity and suppressing probe-induced state leakage represent key directions for further enhancing the readout fidelity in future work.}

\end{document}